\documentclass{aa}  

\usepackage{graphicx}
\usepackage{txfonts}
\usepackage{setspace}

\usepackage{caption}

\usepackage[english]{babel}
\usepackage{graphicx}
\usepackage{subfigure}

\usepackage{txfonts} 
\usepackage{textcomp}
\usepackage{natbib}
\usepackage{amssymb}
\usepackage{graphicx}
\usepackage{tabularx}
\usepackage[english]{babel}
\usepackage[T1]{fontenc}
\usepackage{multirow}

\usepackage{color}

\def\HI{H{\,\small I}}
\def\Htwo{H$_{\,2}$}
\newcommand{\brgamma}{Br$\gamma$}
\newcommand{\sisei}{[Si\,{\small VI}]}

\newcommand{\Htwouno}{H$_{\,2}$ 1-0 S(0,1,2,3)}
\newcommand{\Htwotwo}{H$_{\,2}$ 2-1 S(1,3)}
\newcommand{\Htwoz}{H$_{\,2}$ 1-0 S(0)}
\newcommand{\Htwoi}{H$_{\,2}$ 1-0 S(1)}
\newcommand{\Htwod}{H$_{\,2}$ 1-0 S(2)}
\newcommand{\Htwot}{H$_{\,2}$ 1-0 S(3)}
\newcommand{\Htwodz}{H$_{\,2}$ 2-1 S(1)}
\newcommand{\Htwodt}{H$_{\,2}$ 2-1 S(3)}

\newcommand{\matHI}{\rm H{\hskip 0.02cm\scriptscriptstyle I}}

\newcommand{\kms}{$\,$km$\,$s$^{-1}$}
\newcommand{\ergs}{$\,$erg$\,$s$^{-1}$}
\newcommand{\ergscm}{$\,$erg$\,$s$^{-1}\,$cm$^{-2}$}

\newcommand{\msun}{{${\rm M}_\odot$}}
\newcommand{\msunyr}{{${\rm M}_\odot$ yr$^{-1}$}}

\def\HI{H{\,\small I}}

\def\emph#1{{\sl #1}}
\newcommand{\ltsima} {$\; \buildrel < \over \sim \;$}
\newcommand{\gtsima} {$\; \buildrel > \over \sim \;$}
\newcommand{\lta} {\lower.5ex\hbox{\ltsima}}
\newcommand{\gta} {\lower.5ex\hbox{\gtsima}}

\newcommand{\pks}{\mbox PKS B1718--649}

\begin{document}

   \title{The warm molecular hydrogen of \pks:}

   \subtitle{feeding a newly born radio AGN.}

   \author{F. M. Maccagni\inst{1}$^,$\inst{2}
   \and F. Santoro\inst{1}$^,$\inst{2}
   \and R. Morganti\inst{1}$^,$\inst{2}
   \and T. A. Oosterloo\inst{1}$^,$\inst{2} 
   \and J. B. R. Oonk\inst{2}$^,$\inst{3} 
   \and B. H. C. Emonts\inst{4}}

   \institute{ Kapteyn Astronomical Institute, University of Groningen, Postbus 800, 9700 AV Groningen, The Netherlands
   \and ASTRON, Netherlands Institute for Radio Astronomy, Postbus 2, 7990 AA, Dwingeloo, The Netherlands
   \and Leiden Observatory, Leiden University, Postbus 9513, 2300 RA Leiden, the Netherlands
   \and Centro de Astrobiolog\'ia (INTA-CSIC), Ctra de Torrej\'on a Ajalvir, km 4, 28850 Torrej\'on de Ardoz, Madrid, Spain\\
              \email{maccagni@astro.rug.nl}             }


  \abstract {We present new SINFONI VLT observations of molecular hydrogen (\Htwo) in the central regions ($<2.5$ kpc) of the youngest and closest radio source \pks. We study the distribution of the \Htwo\ traced by the 1-0 S(1) ro-vibrational line, revealing a double disk structure with the kinematics of both disks characterised by rotation. An outer disk ($r>650$ pc) is aligned with  other components of the galaxy (atomic hydrogen, stars, dust), while the inner disk ($r<600$ pc) is perpendicular to it and is polar with respect to the stellar distribution. However, in the innermost $75$ pc, the data show the presence of  \Htwo\ gas  redshifted with respect to the rotating inner disk ($\Delta v\sim +150$\kms) which may trace gas falling into the super massive black hole associated with the central radio source. Along the same line of sight, earlier observations had shown the presence in the central regions of \pks\ of clouds of atomic hydrogen with similar unsettled kinematics. The range of velocities and mass of these unsettled clouds of \HI\ and \Htwo\ suggest they may be actively contributing in fuelling the  central newly-born radio source.}
  

   \maketitle
%

\section{Introduction}
\label{sec:intro}

Active Galactic Nuclei (AGN) are associated with the accretion of material onto the central super-massive black hole (SMBH) of galaxies. The gas surrounding the SMBH must lose angular momentum in order to fall into it so it can trigger and feed an active nucleus. Nevertheless, direct observational evidence of this process is still limited. Statistically, galaxies which have undergone a merger or an interaction event have higher probability to host an AGN \citep{ellison2008,ramos2012,hwang2012,sabater2013}. However, in several objects with signatures of past mergers or accretion, the time-scales associated with these phenomena can be much longer than the age of the AGN~\citep{emonts1,tadhunter2008,schawinski2010,struve2012,maccagni} suggesting that in these galaxies the link between mergers/accretion and AGN is, at most, indirect and other processes must occur to trigger the nuclear activity. Slow secular  processes may help the gas  lose angular momentum on  short timescales ($\sim 10^5-10^8$ years) and form a  dense gas core in the central $100$ pc~\citep{kormendy,wada,Combes2004,combes2011}. However, it is not clear whether these phenomena are efficient in the very innermost regions near the AGN \citep{athan2005,begelman2009}. Thus, other processes taking place on small spatial and temporal scales are expected to be responsible for the direct fuelling onto the AGN~\citep{wada2,king2007,hopkins2010}. Numerical simulations suggest that gravitational and thermal instabilities induce chaotic collisions in the interstellar medium ISM surrounding the SMBH~\citep{soker,gaspari2013,king2008,nayakshin2012,king2015}. This causes small clouds or filaments of gas to lose angular momentum and begin a series of small-scale, randomly oriented accretion events, which then trigger the AGN. In this scenario, the gas deviating from regular rotation is responsible for the chaotic infall of clouds and, consequently, for the accretion onto of the AGN~\citep{gaspari2015}. 

High spatial resolution observations tracing in particular the cold gas in the innermost regions of AGN are needed to investigate these hypothesis. Different type of AGN, i.e.\ Seyfert galaxies~\citep{gallimore,mundell,hicks2009,hicks2013,Combes2014,mezcua}, low-ionisation nuclear emission region galaxies \citep[LINER;][]{garcia-burillo2005,muller2013}, and radio galaxies~\citep{neumayer,dasyra,guillard,morganti2013} are rich in molecular (\Htwo) and atomic hydrogen (\HI) which may represent the fuel reservoir for the nuclear activity. Indeed, the kinematics of at least part of this gas often appears to be unsettled with respect to the regular rotation of the galaxy, suggesting a strong interplay between the nuclear activity and the surrounding environment. On the one hand, it is likely that plasma ejected by the radio source perturbs the neutral and molecular hydrogen~\citep{neumayer,hicks2009,dasyra,guillard,muller2013,mezcua}. On the other hand, it is also possible that this reflects the presence of processes as those described above that can cause the gas to stream towards the SMBH and trigger the nuclear activity~\citep{hopkins2010,Combes2014}. 

Young radio sources in the first stages of their activity~\citep{murgia,fanti} are the best candidates to study the relation between the kinematics of the cold gas and the triggering of the AGN. They are often embedded in a dense gaseous environment where the fuelling of the AGN has just begun and is likely to be still  on-going. Also, among all radio AGN, these sources show relatively often neutral and molecular gas with unsettled kinematics in proximity of the core~\citep{emonts,gereb3,gereb2,curran,guillard2014,allison2015}. 

\begin{figure}
\begin{center}
\includegraphics[trim = 80 0 0 0, clip,width=.52\textwidth]{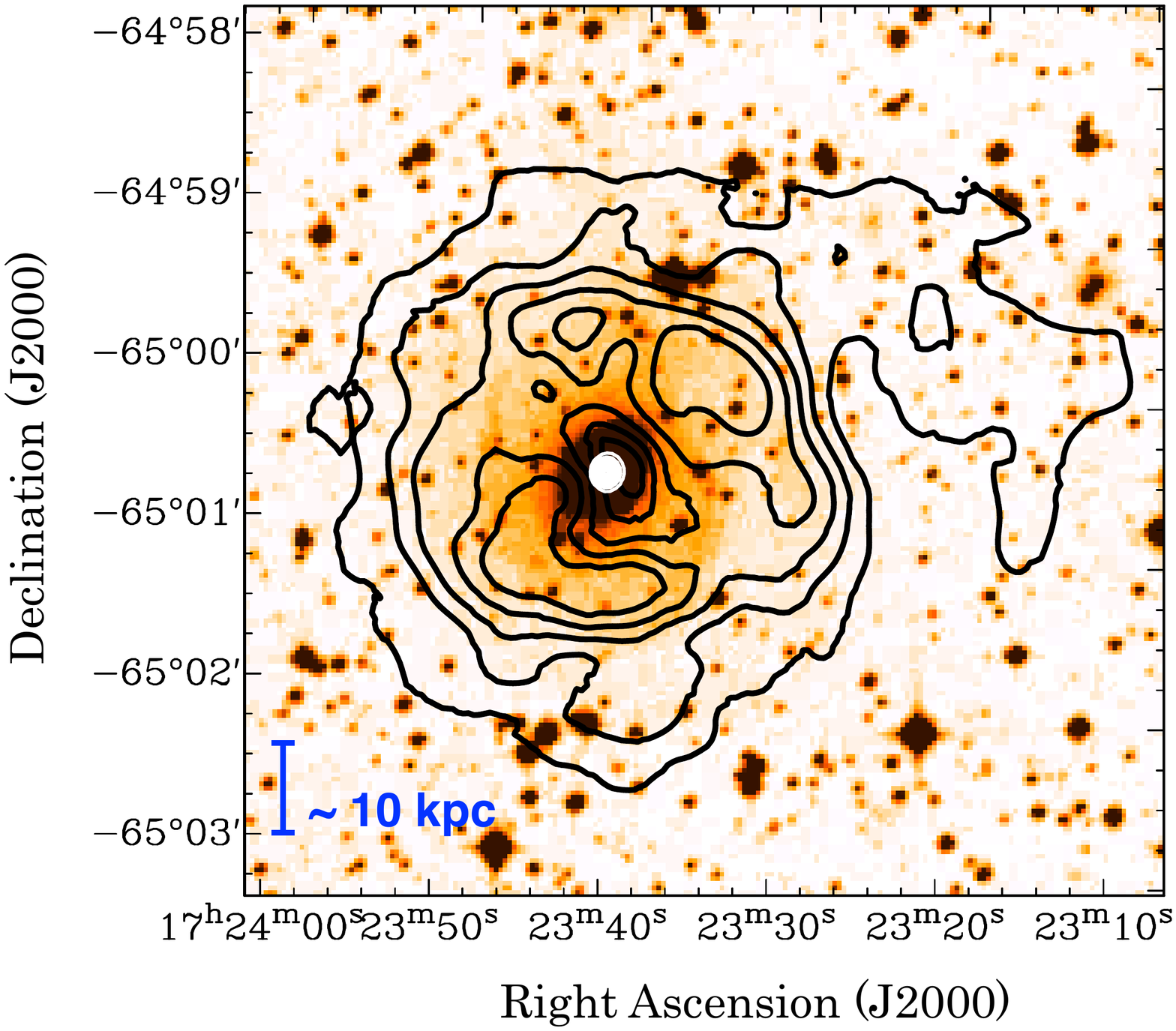}
\caption{$I$-band optical image of \pks, overlaid with the column density contours (black) of the neutral gas. The contour levels range between $7\times10^{19}$ cm$^{-2}$ and $8\times10^{20}$ cm$^{-2}$, with steps of $1.5\times10^{20}$ cm $^{-2}$. The unresolved continuum radio source is marked in white. The \HI\ disk has the shape of an incomplete ring with asymmetries in the N-W and in the S of the disk \citep{maccagni}.}
\label{fig:HI}
\end{center}
\end{figure}

\pks\ is a compact radio source ($r_{\rm radio}\lesssim 2$ pc) with an optically classified LINER AGN~\citep{filippenko} at a distance\footnote{$z=0.014428$; $D_{\rm L}=62.4$ Mpc, 1 arcsec = 0.294 kpc; where $\Lambda$CDM cosmology is assumed, $H_\circ=70$\kms Mpc$^{-1}$, $\Omega_\Lambda=0.7$, $\Omega_{\rm M}=0.3$.} \mbox{of $\sim$62 Mpc}. The estimated age of the radio activity is $\sim$10$^{2\text{-}5}$ years~\citep{tingay1997,giroletti}. \pks\ is morphologically classified as an S0-SABb early-type galaxy embedded in a disk of neutral hydrogen (see Fig.~\ref{fig:HI}) which shows regular rotation out to large radii ($\sim$23 kpc). Given the long time scale for such a regular disk to form, this excludes a merger or a disruptive event being directly responsible for the recent triggering of the central radio source. The accretion onto the SMBH, and the fuelling of the radio activity, could find its origin in a small-scale phenomenon. The detection, in \HI\ absorption, of two separate clouds with kinematics deviating from the rotation of the disk, suggests that a population of clouds may be contributing to feed the AGN in the centre~\citep{maccagni}. The regions close to the radio source have been indirectly probed by the study of the variability of the radio  continuum emission~\citep{tingay2015}, which has been attributed to changes in the free-free absorption due to a clumpy circum-nuclear medium around the radio source. The presence of such a clumpy medium  has been also suggested by optical spectroscopic observations~\citep{filippenko}. The available information on \pks\ suggests that the origin of its newly born radio activity may be found in the kinematics of its circum-nuclear medium.


\begin{table*}
\caption{SINFONI observation specifications}             
\label{table:SINFONI}      
\centering                          
\begin{tabular}{l c }        
\hline\hline                 
Parameter & Value \\    
\hline                        
 Field of view & $8''\times8''$ (2.37$\times$2.37 kpc)\\ 
Pixel size & $0.125''$(37 pc) \\
Spectral resolution (at 2.1 $\mu$m)  & $75$ \kms ($R$=4000) \\
Spectral sampling (at 2.5 $\mu$m) & $36$ \kms \\
Seeing (run 1, run 2, run 3) & $0.4''$ (131 pc); $2.49''$ (735 pc); $0.52''$ (154 pc) \\
\hline                                   
\end{tabular}
\end{table*}

Integral Field Unit (IFU) instruments allow us to analyse the spatial distribution and kinematics of the ISM in the innermost regions of low-redshift AGN. Here we present the results obtained for \pks\ using the Spectrograph for INtegral Field Observations in the Near Infrared (SINFONI) on the VLT. The detection of the molecular hydrogen traced by its ro-vibrational states ($T_{\rm ex}\sim 10^3$ K, \mbox{\Htwouno}) allows us to study its distribution and kinematics and to provide interesting insights on the role of the \Htwo\ in relation to the fuelling of the central radio source.

\section{Observations and data reduction}
\label{sec:data_reduction}

We observed the inner $2.5$ kpc region of \pks\ in the $K$-band ($1.95 - 2.45\,\mu$m), using SINFONI \citep{Eisenhauer2003} mounted on the Very Large Telescope (VLT) UT4.

The observations were performed under seeing-limited conditions during three different nights (May 18-26-29, 2014) in period 93A.  The spectral resolution is $R\sim 4000$ and the plate scale is $0.125''\times0.250''$ pixel$^{-1}$, yielding a field of view of $8''\times8''$. The Full Width Half Maximum (FWHM) of the sky lines is  $6.5\pm0.5$\AA, with a spectral sampling of $2.45$ \AA\ pixel$^{-1}$. Due to bad seeing conditions, we exclude the observations performed on May 26th. 


We perform the data reduction using the official ESO REFLEX workflow for the SINFONI pipeline (version $2.6.8$) and the standard calibration frames provided by ESO. The workflow allows us to derive and apply the corrections for dark subtraction, flat fielding, detector linearity, geometrical distortion, and wavelength calibration to each object and sky frame. Following~\cite{davies2007}, we subtract the sky from the data cubes of the two observing blocks. The typical error on the wavelength calibration is $1.5$ \AA\ ($15$\kms). 

Through IDL routines implemented by \cite{PiquerasLopez2014}, we calibrate the flux of each cube. First, we obtain the atmospheric transmission curve, extracting the spectra of the standard stars with an aperture of $5\sigma$ of the best 2D Gaussian fit of a collapsed cube. Then we normalize these spectra using a black-body profile at the temperature $T_{e}$ corresponding to the spectral type of the observed stars (as tabulated in the 2MASS catalogue, see~\cite{skrutskie}), using Table 5 in \cite{pecaut2013}. We model the stellar hydrogen \brgamma\ absorption line at $2.166\,\mu$m with a Lorentzian profile to determine the sensitivity function for the atmospheric transmission~\citep{bedregal2009}. We convert the star spectra from counts to physical units with a conversion factor extracted using the tabulated $K$ magnitudes in the 2MASS catalogue. We obtain the full-calibrated data cube dividing each spectrum by the sensitivity function and multiplying it by the conversion factor. The typical uncertainty for the flux calibration is $10\%$. We combine the two cubes of the single observing runs into the final one by spatially matching the peaks of the emission of the galaxy. The spatial resolution of the final data cube is equal to the FWHM of a 2D Gaussian we fit to the central region of the collapsed cube: $0.52$ arcseconds.

\begin{figure}
\begin{center}
\includegraphics[trim = 50 0 20 0, clip,width=.49\textwidth]{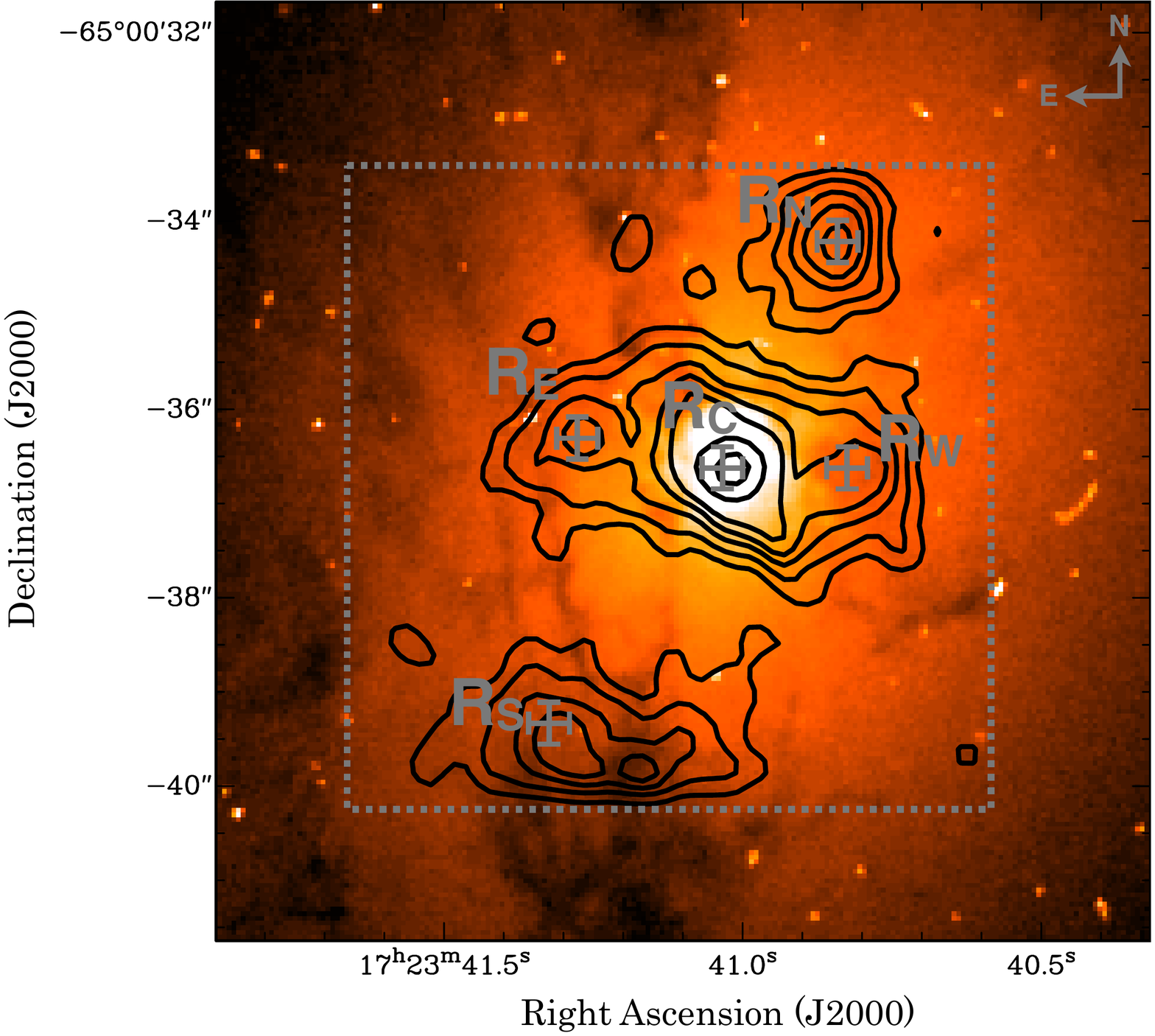}
\caption{Distribution of molecular hydrogen (black contours) overlaid on to the {\em Hubble Space Telescope} WFPC2 image. A dust lane is visible, oriented in the north-south direction. The grey crosses mark the regions where we extracted the spectra to measure the temperature of the \Htwo. The grey dashed square marks the SINFONI field of view.}
\label{fig:dust}
\end{center}
\end{figure}

\begin{figure*}
\begin{center}
\includegraphics[trim = 10 5 10 5, clip,width=.8\textwidth]{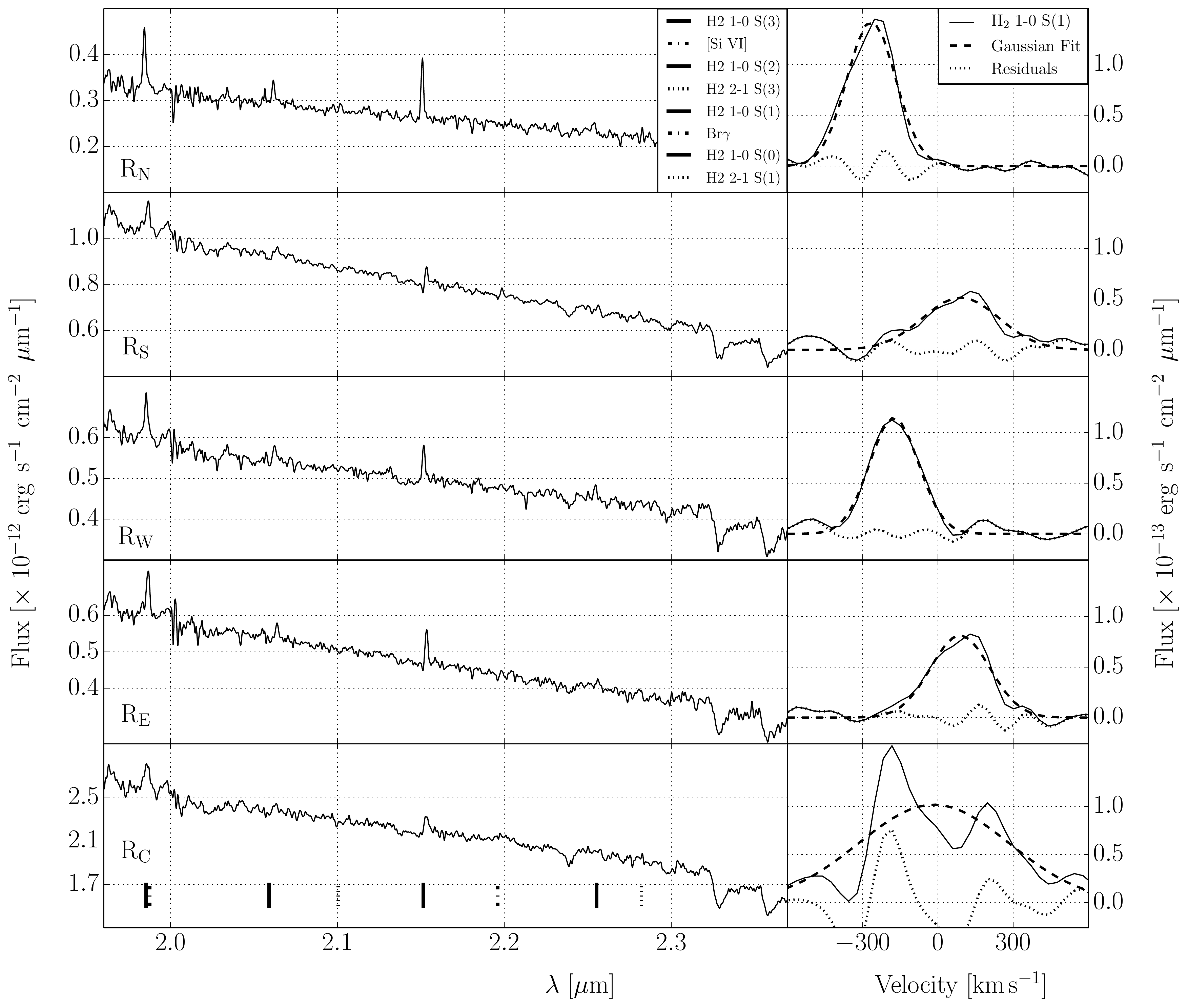}
\caption{Spectra extracted from the five regions (R$_{\rm N}$, R$_{\rm S}$, R$_{\rm W}$, R$_{\rm E}$, R$_{\rm C}$) illustrated in Fig.~\ref{fig:dust}. The right panels show is a zoom-in on the \Htwoi\ line, centred at the systemic velocity of \pks. The dashed line shows the fit with a single Gaussian component, while the dotted line shows the residuals. In the bottom panel, solid lines mark the locations of the \Htwouno\ lines, while dashed and dotted lines show the \Htwotwo\ and \sisei\ and \brgamma\ lines, respectively. }
\label{fig:spectrum}
\end{center}
\end{figure*}

In the final cube, we determine whether a line of the ro-vibrational states of the \Htwo\ is detected considering the spectra extracted over regions of size equal to the spatial resolution of the observations. In Fig.~\ref{fig:spectrum} we show the spectra extracted in five different regions of the \Htwo\ distribution, approximately in the north ($\rm R_{\rm N}$), the south ($\rm R_{\rm S}$), the west ($\rm R_{\rm W}$), the east ($\rm R_{\rm E}$) and in the centre ($\rm R_{\rm C}$) (see Fig.~\ref{fig:dust}). In all five regions, we detect the \Htwoi\ line, the brightest in the spectrum, at $2.14\,\mu$m, as well as the \Htwot\ line, while we do not detect the \Htwo\ 1-0 S(0,2), and the higher excited states of molecular hydrogen, \Htwo\ 2-1 S(1,3). In this case, we determine the $3\sigma$ upper limits assuming a FWHM of the line equal to the one of the \Htwoi\ line. \brgamma\ and \sisei, tracers of high-excitation ionized gas, are also expected in the same spectral range, but lie below the detection limit of these observations.  In Table~\ref{tab:lines}, we list the fluxes and upper limits of the lines for the five different regions.


We focus on the \Htwoi\ line to determine the distribution and the kinematics of the molecular hydrogen and we use the integrated fluxes or the upper limits of the \Htwouno\ lines to determine the temperature of the molecular hydrogen and its mass. We derive the distribution and kinematics of the \Htwoi\ emission line using two independent methods, which provide consistent results. In the first method, we spatially smooth the cube with a Gaussian with FWHM 0.5 arcsec  and we fit a single Gaussian component to the \Htwoi\ line in each pixel of the field of view. We build a mask of the regions of pure line emission, selecting the pixels where the \Htwoi\ line is detected with signal-to-noise ratio of $S/N>5$. We extend the mask to also consider regions neighbouring the ones where the fit is successful. Following this, for each pixel within the mask, we extract the spectrum and we fit the \Htwo\ $1-0$ S(1,3) lines with a single gaussian component. We derive the intensity and velocity fields of the \Htwoi\ line only in the regions where the two lines are detected. 

\begin{table*}
\caption{Line fluxes of the molecular hydrogen gas in five regions of the SINFONI field of view.}             
\label{tab:lines}      
\centering                          
\begin{tabular}{l c c c c c}        
\hline\hline                 
Line [$\lambda_{\rm rest}$] & R$_{\rm N}$ & R$_{\rm S}$ & R$_{\rm W}$ & R$_{\rm E}$ & R$_{\rm C}$ \\    
\hline                        

\Htwot\,\, [$1.95\,\mu$m] & 20.2$\pm$1.29  & 6.88$\pm$1.17 &11.4$\pm$1.52 & 14.1$\pm$1.62 & 8.35$\pm$3.08\\ 

\sisei\,\, [$1.96\,\mu$m]& <3.88 & <3.52 &<4.55 & <4.86 & <8.88\\ 

\Htwod\,\, [$2.03\,\mu$m] & <5.85 & <6.59  &<5.51 & <7.33 & <10.3\\ 

\Htwodt\,\, [$2.07\,\mu$m]& <6.41& <3.64  &<4.89 & <5.90 & <8.04\\ 

\Htwoi\,\, [$2.12\,\mu$m] & 24.2$\pm$0.737 & 11.6$\pm$1.62& 19.1$\pm$0.659 & 16.1$\pm$0.835 & 44.6$\pm$6.15 \\ 

\brgamma\,\, [$2.16\,\mu$m]& <4.01 & <4.01 & <4.56  & <6.19 & <4.55\\ 

\Htwoz\,\, [$2.22\,\mu$m] & <6.39 & <3.61 & <5.63 & <7.54 & <10.2\\ 

\Htwodz\,\, [$2.24\,\mu$m]&<4.73 & <6.81 & <4.59 & <9.09 & <10.4\\ 

\hline                           
\end{tabular}
\tablefoot{ The fluxes are given in units of $\times 10^{-17}$\ergscm. The upper limits indicate the 3-$\sigma$ noise level, measured in the wavelength ranges where we expect to detect the lines. The spectra are extracted in five regions, R$_{\rm N}$, R$_{\rm S}$, R$_{\rm W}$, R$_{\rm E}$, R$_{\rm C}$ of the field of view shown in Fig.~\ref{fig:dust}. Their sizes correspond to the spatial resolution of the observations ($0.52''$).}
\end{table*}

In the second method, we build a cube free of emission-line signal masking out channel-by-channel the regions where the \Htwoi\ line is detected above the 2.5-$\sigma$ level. Next, we smooth the edges of the masked regions to completely exclude any residual emission-line signal. From this cube, we determine the template of the stellar continuum spectrum. We subtract this stellar spectrum from every pixel where the \Htwoi\ line is detected and we obtain a data cube of pure emission-line spectra. We determine the distribution of the \Htwo\ as the zeroth moment map of this cube, summing along the velocity axis all emission above the 3-$\sigma$ level in at least two consecutive pixels. The velocity field corresponds to the first moment map and is centred on the systemic velocity of the \HI\ disk, $v_{\rm sys}=4274$\kms~\citep{maccagni}.

This method is less conservative than the first, but does not rely on the quality of the Gaussian fitting, which may bias the characterisation of the morphology of the \Htwoi\ emission. Interestingly, the total intensity and velocity field determined from the two methods are very similar. In the next section, we use the results of the second method to analyse the kinematics of  the molecular hydrogen in \pks.

The Gaussian fitting of the \Htwoi\ line describes the line across the field of view well, except for in the central $0.52''$, in proximity of the AGN. This is the only region where the fit with a single Gaussian component leaves substantial residuals above $3\sigma$ of the noise (see Fig.~\ref{fig:spectrum} {\em right panels}). The profile is also clearly more asymmetric and the velocity dispersion of the profile is higher than in the other regions of the field of view. This suggests that in the centre, more than one component is needed to fully characterise the kinematics of the \Htwo.  A further analysis of the morphology of the \Htwoi\ emission in the central $75$ pc of \pks\ is given in Section~\ref{subsec:spectrum}.

%
\begin{figure*}
\begin{center}
\includegraphics[trim =30 0 40 0, clip,width=0.49\textwidth]{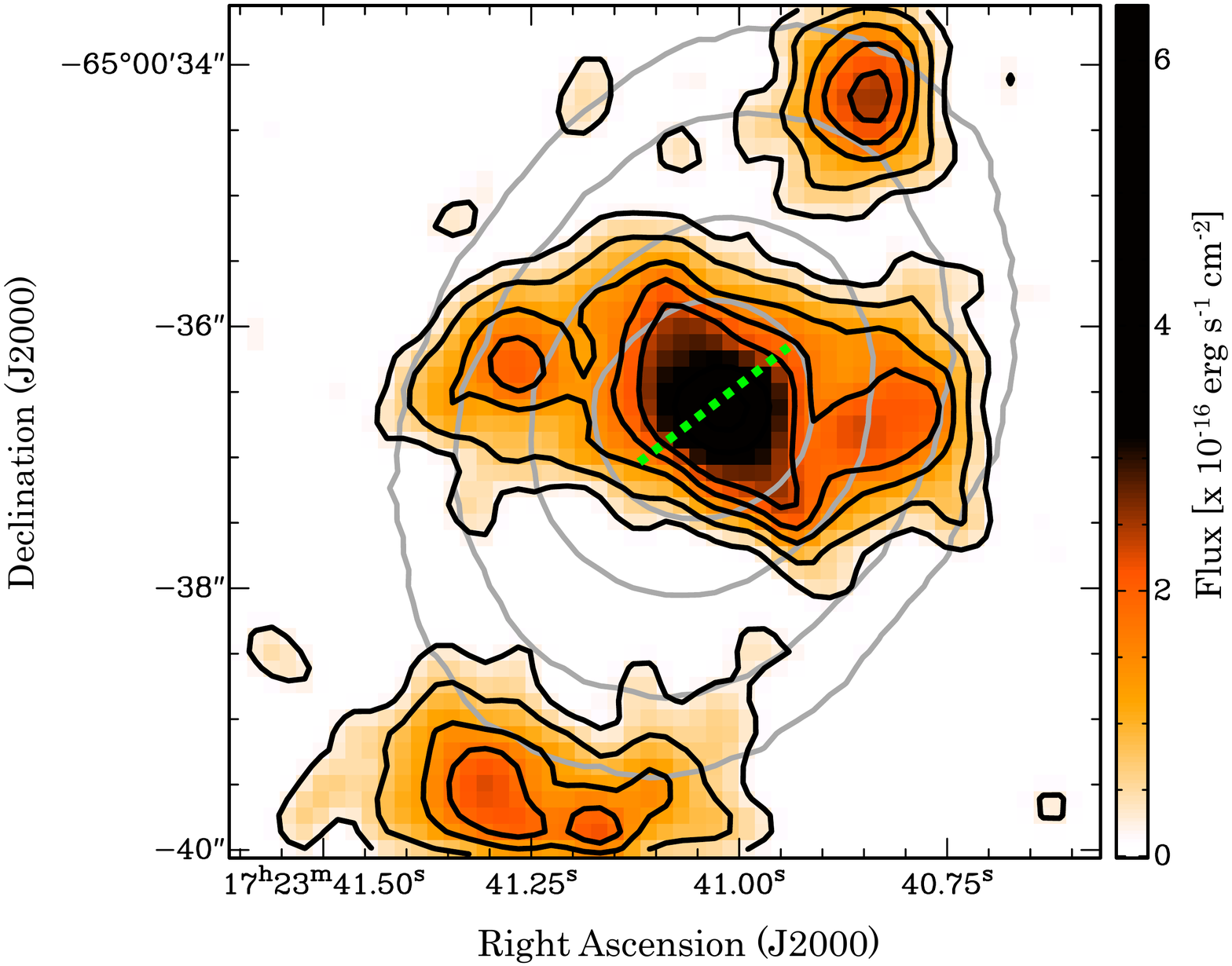}
\includegraphics[trim = 30 0 40 30
, clip,width=0.5\textwidth]{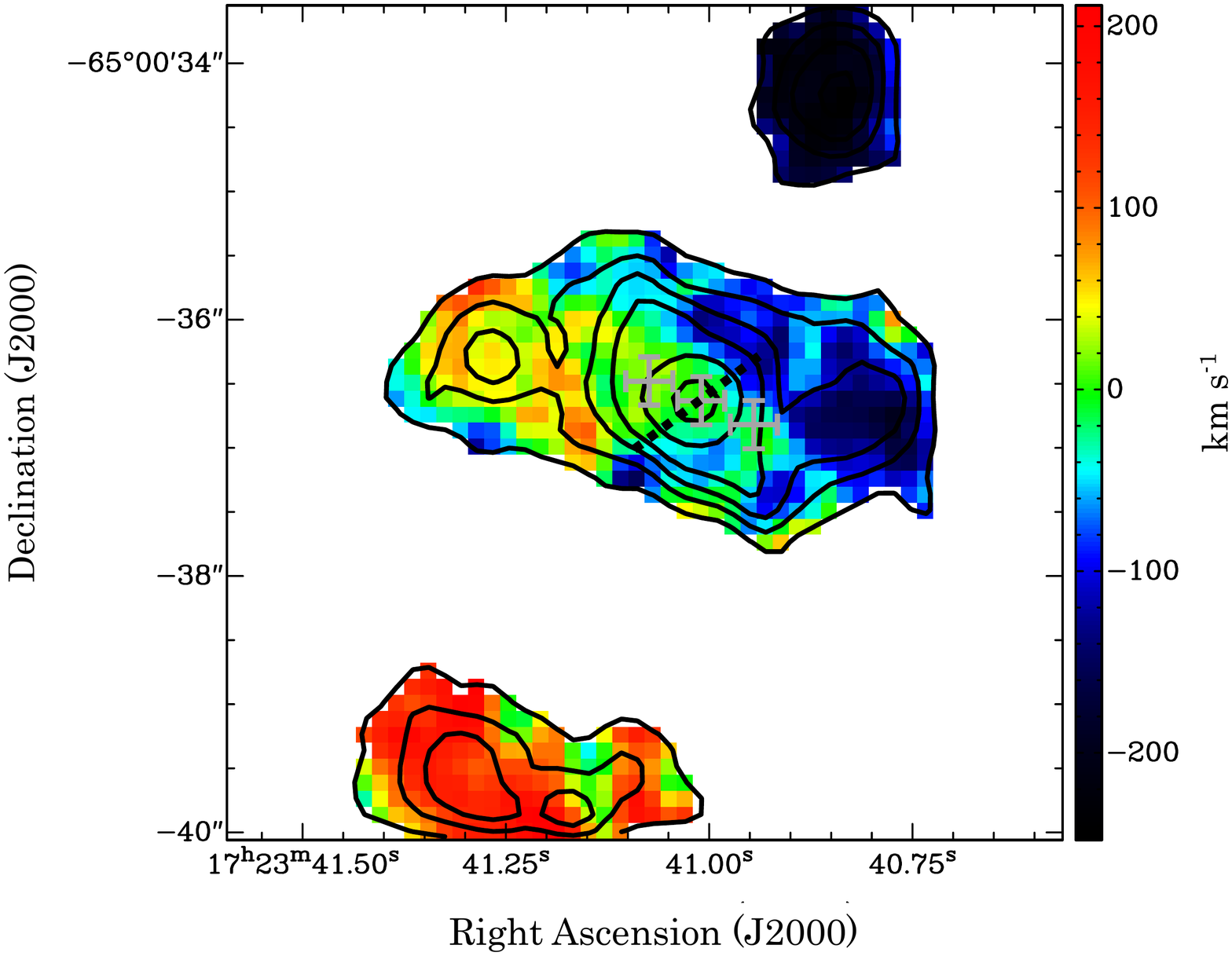}
\caption{{\bf (a)}: Intensity map of the \Htwoi\ line in the inner $2$ kpc of \pks. The position angle of the radio source is shown in {dashed green} (PA$=135^{\circ}$). Intensity contours at 1.5, 3, 5, 7, 9, 12, 15-$\sigma$ are shown in black, starting from 1.5$\sigma = 2.5\times10^{-17}$\ergscm. The isophotes in grey show the distribution of the stellar component. {\bf (b)}: Velocity field of the \Htwoi\ line with the contours of the intensity map overlaid, the position angle of the radio source is shown in black. Velocities are given relative to the systemic value, $v_{\rm sys}=4274$ \kms. The regions marked by crosses are those where we extracted the profiles of Fig.~\ref{fig:spectrum_cont}, (see the text for further details).}
\label{fig:intensity}
\end{center}
\end{figure*}
%

\section{Results}

\subsection{Distribution and kinematics of the molecular hydrogen}
\label{sec:moments}
The intensity map and velocity field of the \Htwoi\ line emission in the central regions of \pks\ are shown in Figs~\ref{fig:intensity}~(a) and (b). At radii $r>650$ pc, the \Htwo\ is assembled in a disk aligned in the N-S direction, the same direction as the \HI\ disk, H$_\alpha$ and the dust lanes at larger radii ($r\sim 8$ kpc)~\citep{keel,maccagni}. The \Htwo\ disk reaches velocities of $\pm200$\kms, similar to the rotational velocities of the large scale \HI\ disk~\citep{maccagni}. From now on, we will refer to it as the `outer disk' of \Htwo. At radii $r<650$ pc, the major axis of the \Htwo\ disk abruptly changes orientation, from approximately north-south (PA$\sim170^\circ$) to east-west (PA$\sim +85^\circ$). Hence, we refer to this as the `inner disk' of \Htwo. The outer disk has asymmetries extending towards the inner disk, possibly suggesting that inner and outer disk are and part of a single, strongly warped structure. From the stellar continuum, we determine the distribution of the stars in the field of view; see the grey isophotes in Fig.~\ref{fig:intensity}~(a). The outer disk is aligned with the stellar component in the N-S direction. Conversely, the inner disk is polar.


In the inner disk, the major axis is aligned in the E-W direction perpendicular to the outer disk. As we move towards the centre, the velocity field suggests that the kinematic minor axis of the disk (green velocities in Fig.~\ref{fig:intensity}~(b)) may change its orientation within $1''$ from the radio source. There, its axis of rotation appears to be aligned with the direction of propagation of the radio jets (dashed line in the Figures). However, given the quality of the data, the presence of this warp should be considered only suggestive.



\subsection{The \Htwoi\ line in the innermost $75$ pc}
\label{subsec:spectrum}

\begin{figure}
\begin{center}
\includegraphics[trim = 0 0 0 0, clip,width=.49\textwidth]{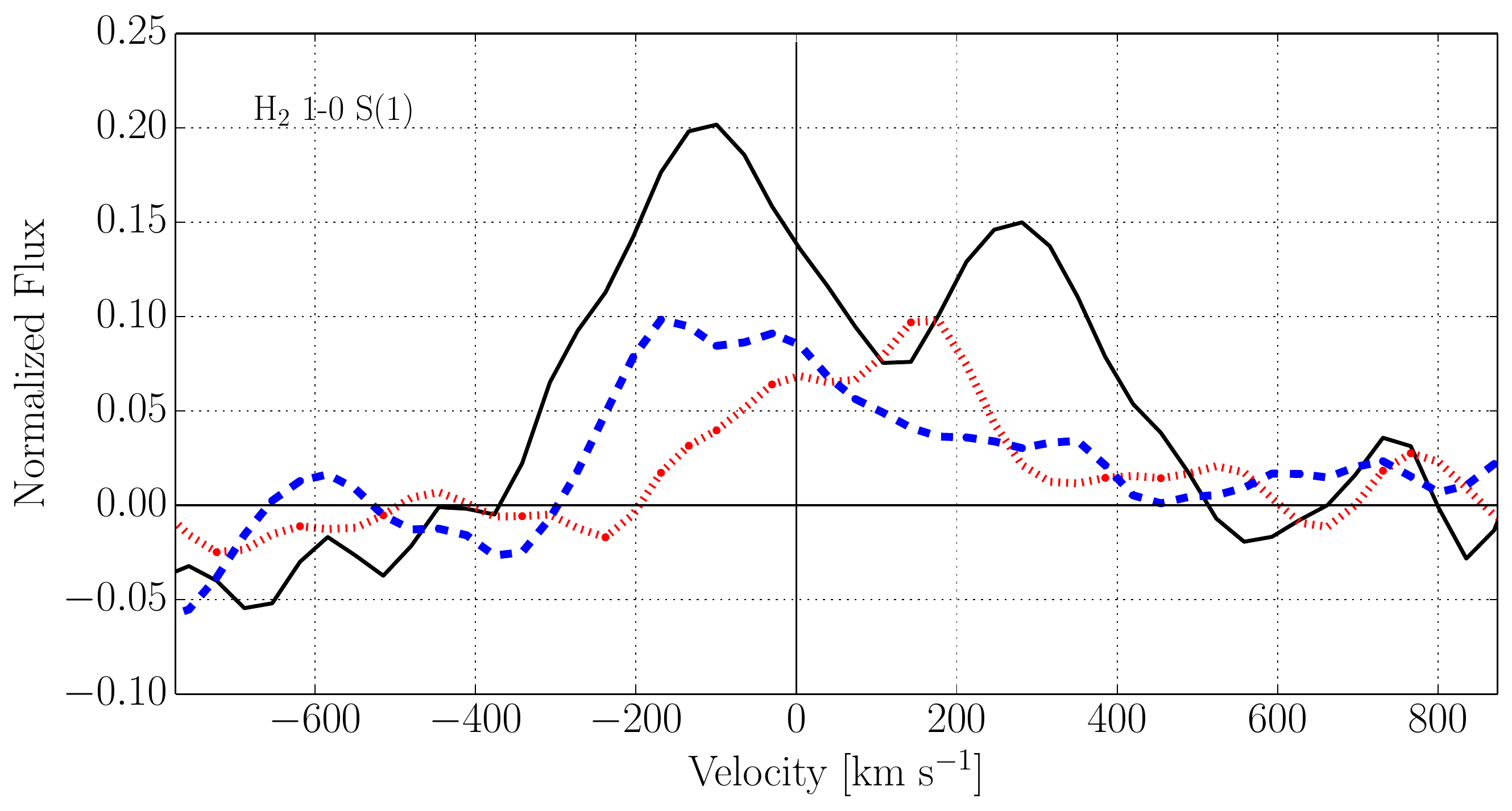}
\caption{Spectra of the \Htwoi\ line, centred at the systemic velocity of \pks. Spectra are extracted along the line of sight to the radio source (black), and on a region on the east (red) and on the west (blue) of the inner disk, at $0.75''$ from the centre. Only the spectrum in front of the radio source appears broader and redshifted.}
\label{fig:spectrum_cont}
\end{center}
\end{figure}

The velocity field in Fig.~\ref{fig:intensity}~(a) suggests that, overall, the disk is dominated by rotation. However, as mentioned above, the very central region ($r<0.25''$)  shows a broader profile compared to the neighbouring regions, suggesting a much larger velocity dispersion in the innermost $\sim75$ pc. In Fig.~\ref{fig:spectrum_cont}, we show the \Htwoi\ line profile extracted from the nucleus ($r\sim0.25''$) and from two adjacent regions on either side of the nucleus (marked by crosses in Fig.~\ref{fig:intensity}~(b)). The line extracted from the central region (C) is shown in black. The dotted red and dashed blue lines, instead, show the \Htwoi\ line from two regions adjacent to the centre, on the east (E) and on the west (W), respectively. All spectra are centred on the systemic velocity of the galaxy. From the figure it is clear that in the centre the \Htwo\ line has an asymmetric profile, with a second component peaking at velocities $>+220$\kms. This component lies outside the range of velocities of the rotation, limited by the flanks of the blue and red lines. The centre of the galaxy ($r<75$ pc) is the only region of the galaxy where the \Htwoi\ line has this feature.

\begin{figure}
\begin{center}
\includegraphics[trim = 0 0 0 0, clip,width=.48\textwidth]{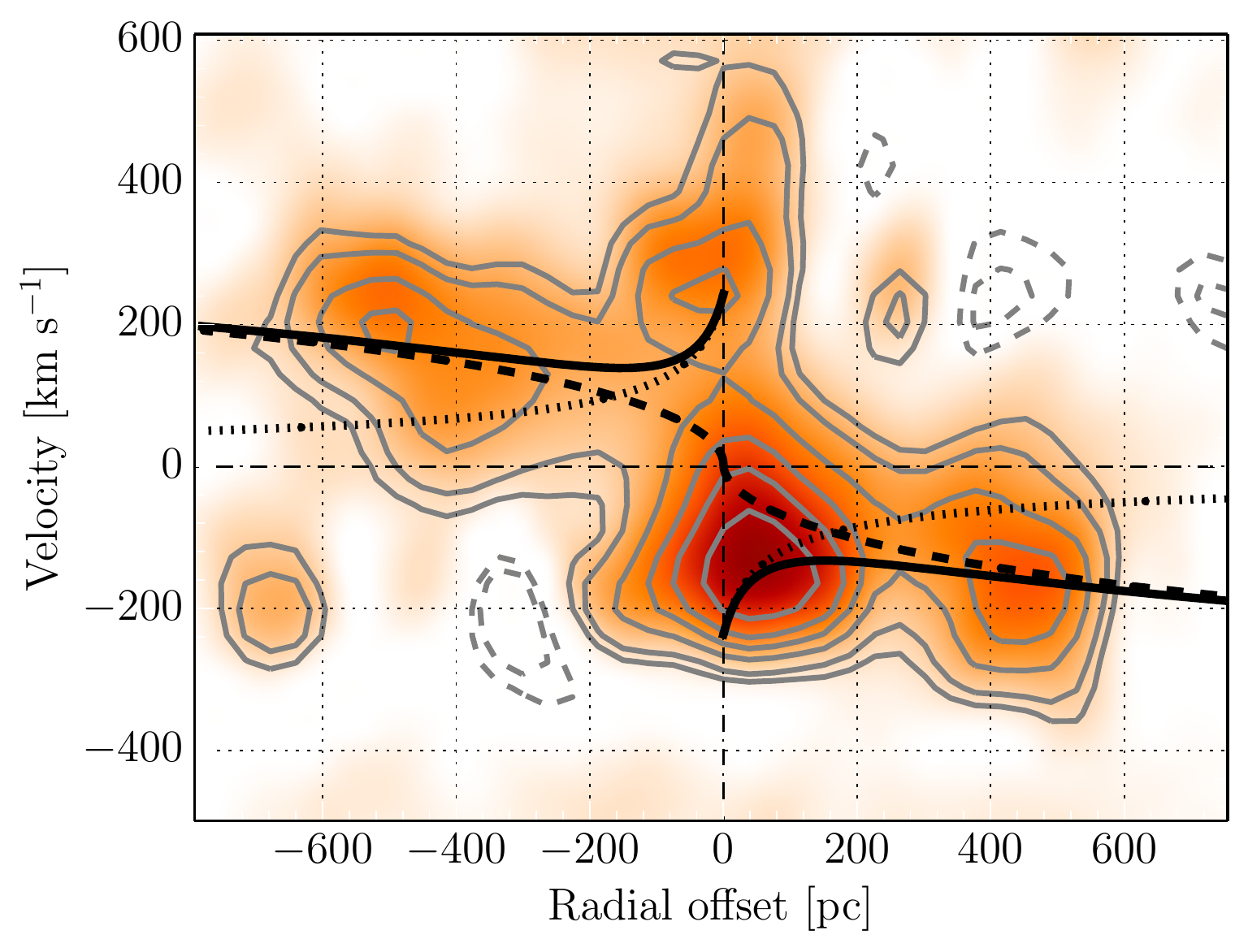}
\caption{Position velocity plot of the \Htwoi\ line extracted along the major axis of the inner disk. Contour levels are  --3, --2, 2, 3, 5, 7,9, and 12-$\sigma$. The black dashed line shows the rotation curve predicted from the stellar photometry, while the fine dashed line shows the contribution of the SMBH to the rotation. The solid line is the total rotation curve derived from the two. In the centre, at velocities $\gtrsim +220$\kms, we identify a component of \Htwo\ deviating from the predicted rotation curve (see Section~\ref{subsec:spectrum} for further details).}
\label{fig:pv_plot}
\end{center}
\end{figure}

This can also be illustrated by  the position-velocity diagram extracted along the major axis of the inner disk {(\bf $PA \sim 85^\circ$)}. Fig.~\ref{fig:pv_plot} shows that the kinematics are characterised by rotation overall, as suggested by the smooth gradient in velocity along the x-axis, symmetric with respect to the centre of the galaxy and with respect to its systemic velocity. Nevertheless, in the centre ($r<75$ pc), the profile appears to be broader and more asymmetric towards redshifted velocities ($v>+200$\kms) than in the rest of the disk. 

Some considerations of the rotation curve of the inner disk of \pks\ allow us to explore this kinematics in more detail. 
\citet{willett} estimate that the mass of the SMBH is $\sim$4$\times10^8$\msun. Assuming that the velocity dispersion of the stars is $\sim$200 \kms, this means that  the SMBH dominates the kinematics of the galaxy out to  $r\sim 45$ pc; while beyond that radius the stellar mass distribution, which is well described by a de Vaucouleurs profile ~\citep{veron}, also contributes. Fig.~\ref{fig:pv_plot} shows a rotation curve based on such a model where we have assumed a total mass of \pks\ of $4 \times 10^{11}$\msun, an effective radius of 9.7 kpc \citep{veron}, circular orbits of rotation, and corrected for the inclination of the inner disk\footnote{The ratio between the minor and major axis of the inner disk, assuming a finite thickness, indicates the disk is oriented approximately edge-on ($i\sim90^\circ$).}. Looking at the central $75$ pc, part of the broad profile can be described by the effect of the SMBH on the gas rotation. However, at redshifted velocities ($v\gtrsim+220$\kms) there is gas extending beyond the velocity range of the rotation curve expected for the mass model: $\Delta v_{\rm uns}\sim+150$\kms. Although our model is fairly qualitative, it suggests gas with anomalous velocities ($\Delta v_{\rm uns}$) exists very close to the SMBH which may be directly involved in its fuelling (see Section~\ref{sec:other_gal} for further details)
.



\subsection{The temperature and mass of the \Htwo}
\label{sec:temperature}
The relative intensity of the \Htwo\ emission lines can be used to infer the temperature and mass of the molecular gas. We estimate the temperature in five different regions within the field of view from the fluxes shown in Table~\ref{tab:lines}. As shown in Fig.~\ref{fig:dust}, we choose two regions in the outer disk (R$_{\rm N}$ and R$_{\rm S}$), two in the inner disk (R$_{\rm N}$ and R$_{\rm S}$) and one in the centre (R$_{\rm C}$). Following~\cite{jaffe2001,wilman2005,oonk2010} (and references therein), assuming that the gas is in local thermal equilibrium, the logarithm of the ratio between the flux of a \Htwo\ line and the flux of the \Htwoi\ line depends linearly on the excitation temperature ($T^{\rm rot}_{\rm ex}$) of the gas itself. The flux ratios of the \Htwo\ 1-0 lines suggest that the inner and outer disk have a temperature between $1100$ K and $1600$ K, while in the centre the temperature is lower $\sim 500$ K. In the R$_{\rm N}$ and R$_{\rm C}$ regions we measure upper limits for the \Htwo\ 1-0 S(0,2) line fluxes which are inconsistent with the local thermal equilibrium (LTE) temperatures derived from the 1-0 S(3) over 1-0 S(1) ratio. Deeper observations are necessary to further investigate these possible deviations from LTE.

The X-ray emission is localized in the innermost $2''$ of the galaxy. Considering that X-rays in the $[0.5-2] \rm\, keV$ energy range could be strongly absorbed, we estimated the lower limit on the luminosity $L_{[0.5-2] \rm\, keV} \gtrsim 10^{41}$\ergs~\citep{maccagni}. This is only one order of magnitude higher than the \Htwo\ luminosity of the {\rm inner disk}, $L_{\rm H_{\,2}} \sim 6.0\times10^{40}$\ergs (where the flux is $5.1\times10^{-14}$\ergscm\, at $D_L\sim62.4$ Mpc). This agrees with the thermal excitation scenario, and suggests that only a small fraction of high-energy photons is required to produce molecular hydrogen emission. The small size of the radio source ($2$ pc) hints that shocks, if present, may excite the warm molecular gas only in the regions right next to the radio jet ($r\ll 75$ pc). Hence, this cannot be the main excitation process of the \Htwo, suggesting that thermal excitation is likely the main responsible mechanism for the warm \Htwo\ emission.

Given the temperature of the molecular hydrogen, we determine the mass of the \Htwo of the inner disk, $M_{\rm H_{\,2}} ({\rm warm})\approx1\times10^4$ \msun, from the flux of the \Htwoi\ line; as shown in \cite{turner,scoville} and~\cite{dale}. From the data cube, we also measure the flux of the unsettled \Htwo\ component in the innermost $75$ pc (see Section~\ref{subsec:spectrum}): $F\sim6.5\times10^{-19}$\ergscm. This corresponds to $M_{\rm H_{\,2}} ({\rm warm})$ $\gtrsim 130$ \msun. 

The amount of warm gas found in the inner disk of \pks\ is in the same range of masses found in the innermost hundreds of parsecs of other LINER galaxies~\citep{muller2013}. Since the \Htwo\ is mainly thermally excited, the \Htwoi\ line may reflect the total mass of the cold molecular component, i.e. the \Htwo\ in its ground state ($T_{\rm exc}\sim100$~K) commonly traced by the CO lines and the \Htwo\ 0-0 S(0,...,7) rotational lines. Within one order of magnitude, the mass of the cold \Htwo\ can be estimated from the mass of the warm \Htwo.  We find $M_{\rm H_{\,2}} ({\rm cold}) \approx 2\times 10^9$ \msun\ for the inner disk, and $M_{\rm H_{\,2}}\ ({\rm cold})$ $\gtrsim 5 \times 10^7$ \msun\ for the unsettled \Htwo\ in the central $75$ pc. Where we use the relation found for a sample of galaxies with similar morphological classification as \pks,~\citep{muller2006,dale,mazzalay,emonts2014},.



\begin{table}[tbh]
 \centering
 \renewcommand{\tablename}{Tab.}
\caption{Main properties of the molecular hydrogen in the innermost regions of \pks.}
\begin{tabular}{l c c}
\hline \hline
Parameter & Inner disk  & Deviating component  \\
 \hline
Radius [pc]& <650 & < 75 \\ 
Flux [\ergscm] & $5.1\times10^{-14}$ &  $1.5\times10^{-15}$\\
Luminosity [\ergs] & $3.5\times10^{40}$ & $5.7\times10^{38}$\\
Temperature [K] & $\sim1100$ & $\sim900$ \\
Mass \Htwo\ (warm) [\msun] & $ \sim1.5\times 10^{4}$ & $\gtrsim 130$ \\
Mass \Htwo\ (cold ) [\msun] & $\sim2\times 10^{9}$ & $\sim 5\times10^7$ \\
\hline
\end{tabular}
\label{tab:Htwo}
\end{table}

\section{Relating the kinematics of the gas to the radio nuclear activity}
\label{sec:other_gal}

The SINFONI observations in the innermost kilo-parsec of \pks\ reveal two disks of molecular hydrogen. The outer disk ($r>650$ pc), oriented in the N-S direction, follows the rotation of the stars and of the other gaseous components of the galaxy. The inner disk ($r\lesssim 600$ pc) is oriented E-W with kinematics overall characterised by rotation. In Section~\ref{subsec:spectrum}, we showed that in the innermost $75$ pc of \pks, the \Htwoi\ line is brightest and asymmetric, suggesting the presence of a second component of \Htwo\ with unsettled kinematics deviating from the rotation with redshifted velocities $\Delta v_{\rm uns} \sim +150$\kms. 

The \Htwo\ is not the only gaseous component with unsettled kinematics near the radio source. Along the same line of sight (and in particular only in front of the central $2$ pc of the radio source) the \HI\ also shows kinematics deviating from regular rotation~\citep{maccagni}. Two separate absorption lines, with opposite velocities with respect to the systemic value, suggest the presence of small clouds of cold gas close to the AGN that deviate from the rotation of the other components of the galaxy. From the separation between the two lines, we estimate that these clouds have unsettled velocities of  $v_{\rm uns}(\matHI)\sim 100$\kms. Given that the distribution of the \Htwo\ in the inner disk is not homogeneous, it is reasonable to assume that the \HI\ clouds are located in the same region of the \Htwo\ with unsettled kinematics, i.e. in the innermost $75$ pc of the galaxy. The presence of a clumpy multiphase environment around the radio source is also suggested by the variability of its radio-continuum~\citep{tingay2015}, which has been attributed to changing conditions in the free-free absorption in a surrounding clumpy cold medium. Moreover, optical spectroscopic observations also suggest the presence of a clumpy circum-nuclear medium~\citep{filippenko}. A similar distribution of \Htwo, with increasing velocity dispersion in the central $\sim$100 pc, has been detected in a number of different AGN and Seyfert galaxies~
\citep{hicks2009,hicks2013,davies2014,guillard,muller2013,mazzalay,mezcua}. 


\pks\ is thus a newly born compact radio AGN surrounded by a rotating clumpy multi-phase circum-nuclear disk, where we measure deviations from rotation in the \HI\ and the \Htwo\ only in the innermost $75$ pc. While the \HI, because it is detected in absorption, must be located in front of the radio source, the \Htwo\ is detected in emission and can be located either in front and/or behind the radio source. Hence, in principle, the redshifted velocities of the unsettled gas could correspont to either an infall or an outflow. It is difficult to disentangle this from the available data. However, given the properties of this AGN we note it is unlikely that it is driving an outflow. The small scale of the radio source ($2$ pc) and its low jet-power ($P_j\lesssim 2.3\times 10^{43}$\ergs) would exclude a jet-driven outflow. Since \pks\ is a LINER galaxy, the radiation from the optical AGN is also limited ($P_{\rm rad} \lesssim 8 \times 10^{43}$\ergs) and an outflow is not likely to occur on energetic grounds. These considerations make it plausible to assume that the redshifted unsettled velocities of the \Htwo\ are connected to gas falling into the AGN and perhaps being responsible for its fuelling. \pks\ does not show traces of previous periods of radio activity which could have perturbed the gas\footnote{since the 1.4 GHz continuum flux over $\sim 4$ pc$^2$ (beam of the VLBI observations,\cite{tingay2002}) is the same as over $\sim 64$ kpc$^2$ (resolution of the ATCA observations, \cite{maccagni}).}. Hence, it is likely that, when the radio source was triggered, these clouds with unsettled kinematics ($\Delta v_{\rm uns}\sim +150$\kms) were already present in the innermost $75$ pc of the circum-nuclear disk. The double disk structure of the \Htwo, and the large-scale strongly warped \HI\ disk, suggest that the gas in \pks\ is still settling in the gravitational potential and   that stellar torques are acting on the  gas to align  into a stable configuration. These torques may strip gas clouds from the inner two-disk configuration so that the clouds subsequently become unsettled and fall toward the SMBH.


Simulations of black hole accretion in rotating environments~\citep{king2008,nayakshin2012,gaspari2013,gaspari2015} have suggested that, because of the local instabilities of the medium, chaotic collisions between clouds, cold filaments and the clumpy circum-nuclear disk may unsettle the kinematics of the gas and promote the cancellation of angular momentum. This may lead to the triggering of accretion into the SMBH. If the velocity dispersion does not exceed the rotational velocity, the accretion rates onto the AGN are predicted to be $\lesssim 0.1$ \msunyr. This scenario could be an alternative explanation for the disturbed kinematics observed in the \HI\ and the \Htwo\ of \pks, where the deviations from rotation ($\Delta v_{\rm uns}\sim+150$\kms) are on the same order of magnitude as the rotational velocity ($v_{\rm rot}\sim 220$\kms). This scenario predicts inefficient accretion onto the AGN, which is also suggested by the radio power and by the LINER nature of \pks.

In low-efficiency radio AGN, the radio power may set a constraint on the accretion rate onto the SMBH~\citep{allen,balmaverde}. In \pks\, this is equal to $\dot{M}\lesssim 10^{-2}$ \msunyr. In \cite{maccagni}, we derived a limit for the contribution of the \HI\ on the accretion using a very uncertain distance of the clouds from the nucleus due to the large beam of the \HI\ observations. The \Htwo\ emission allows us to constrain this distance to $\lesssim 75$ pc, hence we determine the accretion rate of the \HI\ clouds and of the \Htwo\ with unsettled kinematics, and investigate whether this could sustain the radio activity. Assuming the velocities deviating from rotation are equal to the in-fall velocity into the black hole ($v\sim +150$ \kms) and assuming a distance of the clouds from the SMBH $r\lesssim75$ pc, we determine a typical timescale of accretion of these components to be $t_{\rm accretion}\sim 6.9\times 10^5$ years.  
The mass of the \HI\ clouds is constrained by the column density of the absorption lines; assuming these are located within $\sim75$ pc from the radio source, we determine $M_{\rm H\,\small I}\sim3.5\times 10^2$ \msun. From this, it follows $\dot{M}_{\rm H\,\small I}\sim 10^{-4}$ \msunyr, which is insufficient, alone, to sustain the radio activity. In the innermost $75$ pc, the warm molecular hydrogen with unsettled kinematics has a mass of $M_{\rm H_2} ({\rm warm})\lesssim130$ \msun, which also gives an accretion rate of $\dot{M}_{\rm H_2}\sim 10^{-4}$ \msunyr. If some of the cold \Htwo\ ($T_{\rm ex}\sim10^2$ K, see Section~\ref{sec:temperature}) is also involved in feeding the AGN,  we may obtain an accretion rate sufficient to power such radio source.

\pks\ has some interesting features in common with the nearest radio galaxy Centaurus A. Like \pks, Centaurus A has a young radio core surrounded by a circum-nuclear rotating disk of \Htwo\ that is embedded in a large-scale \HI\ disk~\citep{struve2010}. Centaurus A  also shows a brighter and asymmetric \Htwo\ line profile, in the innermost $200$ pc. This can be explained by gas streaming down into the AGN~\citep{neumayer}. \pks\ appears to be another example where we witness the fuelling of a radio-loud AGN. We plan to furhter investigate this with future observations.

\section{Conclusions}
\label{sec:conclusions}

Our SINFONI $[1.95-2.45]\,\mu$m observations of the innermost $8''\times 8''$ of \pks\ have shown the presence of molecular hydrogen assembled into two orthogonal disks. The outer ($r>650$ pc) disk of \Htwo\ is oriented along the north-south direction aligned with the stellar distribution and of which the kinematics connects smoothly to that of the large-scale \HI\ disk. At radii $r<650$ pc, the \Htwo\ is assembled in an inner circum-nuclear disk, aligned in the east-west direction and polar with respect to the stars. The kinematics of the disks is characterised by rotation with velocities of about $220$\kms. Assuming thermal equilibrium within the disk, we determine the temperature of the \Htwo\ to be $T_{\rm ex}\sim1100$ K and its mass $M_{\rm H_{\,2}}\,({\rm warm})\approx1\times10^4$ \msun, which may trace up to $\sim 2 \times 10^9$ \msun\ of cold molecular hydrogen. 

The kinematics of the inner disk of \Htwo\ is characterised by rotation due to the combination of the stellar distribution and  the SMBH (see Section~\ref{subsec:spectrum}). In proximity of the radio source, at radii $r<75$ pc, we detect \Htwo\ deviating from such rotation. In the innermost $75$ pc the \Htwo\ has unsettled kinematics in the range $\Delta v_{\rm uns}\sim +150$\kms. This component of warm \Htwo\ has a mass $\sim 130$~\msun, which may trace $\lesssim 5 \times 10^7$ \msun\ of cold molecular hydrogen. The \HI\ clouds detected in absorption against the compact radio core by \citet{maccagni} have similar velocities deviating from rotation and could be located in the same region close to the radio source. These observations, along with the information collected from the variability of the radio continuum~\citep{tingay2015} and the line ratios of the optical forbidden lines~\citep{filippenko}, suggest that the circum-nuclear ISM is clumpy and may represent the fuel reservoir of the radio source. The mass traced by the \HI\ clouds and by the warm \Htwo\ alone is insufficient to fuel the AGN to power the radio jets. Instead, the mass of total cold \Htwo\ ($T_{\rm ex}\sim10^2$ K) traced by the warm unsettled \Htwo\ in the innermost $75$ pc could fuel the radio source at the required accretion rate. Given the low power of the AGN, inefficient accretion is most likely to occur in \pks. Given the double disk structure of the \Htwo, which is part of the larger ($r\sim 23$ kpc) \HI\ disk, the gas configuration could be caused by the stellar torques acting on the gas to align  into a stable configuration and which may give rise to small clouds with unsettled kinematics. The small clouds of \HI\ and \Htwo\ with unsettled velocities of $\sim 150$\kms, that we detect, could be falling into the AGN, contributing to the fuelling of the radio source.

.

\begin{acknowledgements}
The research leading to these results has received funding from the European Research Council under the European Union's Seventh Framework Programme (FP/2007-2013) / ERC Advanced Grant RADIOLIFE-320745. BE acknowledges funding by the European Union 7th Framework Programme (FP7-PEOPLE-2013-IEF) grant 624351. The authors wish to thank J. Piqueras L\'opez for the help in the data reduction and the development of the IDL routines. SINFONI is an adaptive optics assisted near-infrared integral field spectrometer for the ESO VLT. The observations presented in this paper have been taken at the La Silla-Paranal Observatory under programme 093.B-0458(A). 
\end{acknowledgements}


\bibliographystyle{aa}
\bibliography{biba}

\end{document}